\documentclass[aps,twocolumn,prl,letterpaper]{revtex4}
\usepackage[pdftex]{graphicx}% include figures
\usepackage{amssymb} % AMS math symbol
\usepackage{url}     % handle url
\usepackage{bm}      % bold math
\usepackage[normalem]{ulem}    % strikethrough font.
\usepackage{color}
\usepackage{cancel}
\usepackage{float}
\begin{document}
\newcommand{\ohm}{\ensuremath{\,\Omega}}
\newcommand{\Ef}{\ensuremath{E_\mathrm{F}}}
\newcommand{\Bf}{\ensuremath{B_\mathrm{f}}}
\newcommand{\Bc}{\ensuremath{B_\mathrm{c}}}
\newcommand{\kf}{\ensuremath{k_\mathrm{F}}}
\newcommand{\vf}{\ensuremath{v_\mathrm{F}}}
\newcommand{\rc}{\ensuremath{r_\mathrm{c}}}
\newcommand{\rcbg}{\ensuremath{r_\mathrm{c}^{\mathrm{BG}}}}
\newcommand{\rctg}{\ensuremath{r_\mathrm{c}^{\mathrm{TG}}}}
\newcommand{\Vtg}{\ensuremath{V_\mathrm{TG}}}
\newcommand{\Vbg}{\ensuremath{V_\mathrm{BG}}}

\newcommand{\ctg}{\ensuremath{C_\mathrm{TG}}}
\newcommand{\cbg}{\ensuremath{C_\mathrm{BG}}}

\newcommand{\ntg}{\ensuremath{n_\mathrm{TG}}}
\newcommand{\nbg}{\ensuremath{n_\mathrm{BG}}}

\newcommand{\Leff}{\ensuremath{L_\mathrm{eff}}}

\newcommand{\pn}{\ensuremath{p\textrm{--}n}}
\newcommand{\dGVtg}{\ensuremath{\mathrm{d}G/\mathrm{d}V_{\mathrm{TG}}}}
\newcommand{\cred}{\color{red}}

\setlength{\pdfpageheight}{\paperheight}
\setlength{\pdfpagewidth}{\paperwidth}

\title{Conductance oscillations induced by ballistic snake states in a graphene heterojunction}
\author{Thiti  Taychatanapat$^{1,2}$}
\author{Jun You Tan$^{1,2}$}
\author{Yuting Yeo$^{1,2}$}
\author{Kenji Watanabe$^{3}$}
\author{Takashi Taniguchi$^{3}$}
\author{Barbaros \"{O}zyilmaz$^{1,2,4}$}
\affiliation{$^{1}$Graphene Research Centre, National University of Singapore, 117542, Singapore}
\affiliation{$^{2}$Department of Physics, National University of Singapore, 117542, Singapore}
\affiliation{$^{3}$National Institute for Materials Science, Namiki 1-1, Tsukuba, Ibaraki 305-0044, Japan}
\affiliation{$^{4}$NanoCore, National University of Singapore, 117576, Singapore}
\date{\today}

\begin{abstract}
The realization of \textit{p}--\textit{n} junctions in graphene, combined with the gapless and chiral nature of its massless Dirac fermions has led to the observation of many intriguing phenomena such as quantum Hall effect in bipolar regime, Klein tunneling, and Fabry-P\'{e}rot interferences all of which involve electronic transport across \textit{p}--\textit{n} junctions. Ballistic snake states propagating along the \textit{p}--\textit{n} junctions have been predicted to induce conductance oscillations, manifesting their twisting nature. However, transport studies along \textit{p}--\textit{n} junctions have so far only been performed in low mobility devices. Here, we report the observation of conductance oscillations due to ballistic snake states along a \textit{p}--\textit{n} interface in high quality graphene encapsulated by hexagonal boron nitride. These snake states are exceptionally robust as they can propagate over $12$~$\mu$m, limited only by the size of our sample, and survive up to at least $120$~K. The ability to guide carriers over a long distance provide a crucial building block for graphene-based electron optics.
\end{abstract}

\maketitle

Snake states, a counterpart of skipping orbit states, were first proposed to exist along the boundary between positive and negative magnetic field\cite{Vilms_inhomogeneous_B,Muller_snake_B_0,Oroszlany_snake_B_1,Ghosh_snake_B_2,Davies_snake_B_gradient}. However, the difficulty to achieve a magnetic field gradient experimentally\cite{Nogaret_B_gradient} has led to a new proposal based on a $\pn$ junction in graphene\cite{Carmier_snake_pn_2010,Beenakker_review} where, instead of inverting the magnetic field polarity, we interconvert carrier types between hole and electron across the junction, generating snake-like trajectories at finite magnetic field. An increase in conductance along the $\pn$ interface of graphene devices on SiO$_2$ at finite magnetic field has been observed and attributed to snake states\cite{Williams_snake_states}. It has been theoretically predicted that, in a higher quality device, these snake states, even though propagating along the $\pn$ interface, can manifest themselves through oscillatory conductance across the interface as a result of alternating end points of the snake states\cite{Carmier_snake_pn_2010,Chen_snake_current_oscil,Patel_caustics,Milovanovic_snake_Hall_bar,Milovanovic_MEF_PN}. The high-quality $\pn$ junction in graphene, only feasible recently following the development of graphene heterostructures\cite{Dean_BN,Mayorov_BendResistance}, provides a perfect platform for snake states due to its gapless and chiral nature which leads to a  transparent $\pn$ junction at normal incident angle, owing to Klein tunneling\cite{Katsnelson_Klein,Cheianov_transmission,Shytov_Klein_FP,Young_FP_oscil}.

In this work, we investigate conductance oscillations arising from ballistic snake states by employing a top-gated graphene encapsulated by hexagonal boron nitride (hBN) in order to access the ballistic regime\cite{Dean_BN,Mayorov_BendResistance,Thiti_focusing}. The observed conductance oscillations are present only at finite magnetic field and in bipolar regime. The density and magnetic field dependence of these oscillations supports a theory of snake states propagating along the $\pn$ junction. The oscillations persist up to at least $120$~K, showing the robustness of the snake states.

%%%%%%%%%%%%%%%%%%%%%%      Figure 1     %%%%%%%%%%%%%%%%%%%%%%%%%%%%%%%%%%%%%%
%%%%%%%%%%%%%%%%%%%%%%%%%%%%%%%%%%%%%%%%%%%%%%%%%%%%%%%%%%%%%%%%%%%%%%%%%%%%%%%
\begin{figure*}
\begin{center}
\includegraphics{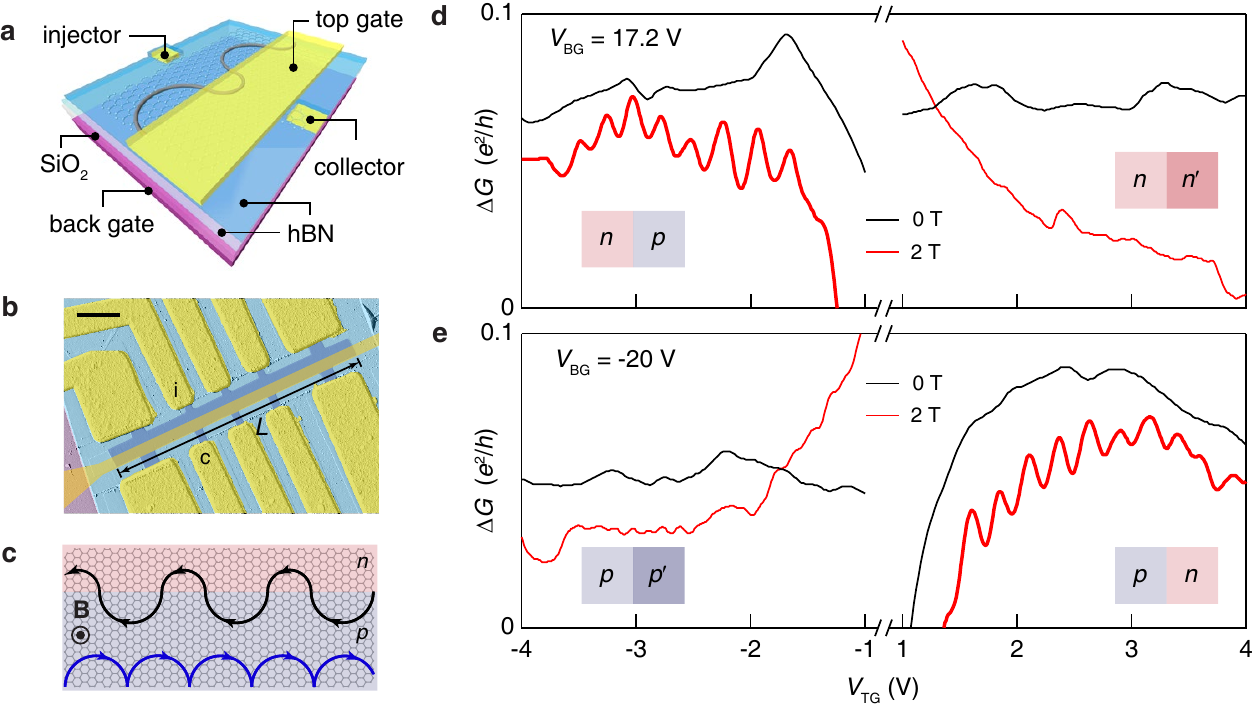}
\caption{{\bf Device schematic and conductance measurements.} ({\bf a}) Schematic diagram of a graphene heterojunction device. Graphene is encapsulated by hBN and top gate and back gate are used to create a $p$-$n$ junction. ({\bf b}) False-colour AFM image of a device after mechanical cleaning with an AFM tip. The contacts labeled by i and c are used for an injector and collector with the width of $\sim$$300$~nm.  The width and length of graphene are $900$~nm and $12.2$~$\mu$m respectively. A top gate (orange) is overlaid whose width is $500$~nm. The edge of the top gate is aligned with the graphene edge. The scale bar is 2 $\mu m$. ({\bf c}) Classical trajectories of skipping orbits (blue) and snake states (black). Magnetic field $\mathbf{B} = B\mathbf{z}$ is applied normal to graphene surface where $\mathbf{z}$ is the unit vector normal to the surface. ({\bf d}, {\bf e}) Conductance versus $\Vtg$ at $\Vbg = 17.2$, $-20$~V respectively. Black and red curves represent conductances at $0$ and $2$~T respectively. Conductance oscillations (thick red line) can be seen only at finite field in the bipolar regime. Curves are offset for clarity.}
\label{F:Fig1}
\end{center}
\end{figure*}
%%%%%%%%%%%%%%%%%%%%%%%%%%%%%%%%%%%%%%%%%%%%%%%%%%%%%%%%%%%%%%%%%%%%%%%%%%%%%%%
%%%%%%%%%%%%%%%%%%%%%%%%%%%%%%%%%%%%%%%%%%%%%%%%%%%%%%%%%%%%%%%%%%%%%%%%%%%%%%%

Figures 1a and b show a schematic diagram and an atomic force microscopy (AFM) image of our device. Graphene is transferred onto a $32$-nm-thick hBN and etched into a Hall bar geometry whose width and length is $900$~nm and $12.2$~$\mu$m respectively. The device is then encapsulated by an $8$-nm-thick hBN. A top gate is defined onto the device, parallel to the Hall bar (see Methods for fabrication). The thin top hBN flake allows us to create a sharp $\pn$ interface which increases the transmission probability across the $\pn$ junction\cite{Cheianov_transmission}. We measure conductance ($G$) using a two probe method in which carriers are injected into graphene through an injector (i) and collected at a collector (c).  Back gate (BG) and top gate (TG) allow us to control densities in back-gated and top-gated regions (BGR and TGR) independently.

Under a finite magnetic field normal to graphene surface ($\mathbf{B} = B\mathbf{z}$), carriers, injected into graphene, will undergo a skipping orbit motion along the edge (Fig.~1c, blue curve). After reaching a $\pn$ interface, some of the carriers will convert into snake states (Fig.~1c, black curve), propagating along the interface and inducing conductance oscillations. Figure 1d shows conductance as a function of top gate voltage ($\Vtg$) at $\Vbg = 17.2$~V, when the BGR is $n$-doped. As we sweep $\Vtg$, carrier density in the TGR can be tuned from $p$-type doping to $n$-type doping, creating $n$-$p$ or $n$-$n'$ junctions respectively. At $B = 0$~T (black line), the conductance shows no oscillations as a function of $\Vtg$ in both unipolar and bipolar regime. However, at $B = 2$~T (red line), clear oscillations emerge in the bipolar regime ($ -4$~V~$\leq \Vtg \leq -1$~V, thick red line) but is absent in the unipolar regime ($1$~V~$\leq \Vtg \leq 4$~V). When BGR is $p$-doped (Fig.~1e), similar oscillations can also be observed at finite field in the bipolar regime ($1$~V~$\leq \Vtg \leq 4$~V).

The oscillations we observe here only exist in the bipolar regime which allows us to rule out Shubnikov-de Haas (SdH) oscillations as a main mechanism. Similar oscillations in the bipolar regime are also observed in graphene heterojunctions as a result of Fabry-P\'{e}rot (FP)  interferences\cite{Young_FP_oscil}. However, one distinct difference is that our oscillations are absent at zero magnetic field while FP oscillations can be observed at zero field. This dismisses the possibility that our oscillations originate from the FP interferences.

In order to gain more insight into the nature of these oscillations, we measure conductance as a function of $\Vbg$ and $\Vtg$. To eliminate the background conductance, the numerical derivative of $G$ with respect to $\Vtg$ is analyzed (see Supplementary Fig. 1 for data before numerical derivatives).   At $B = 0$~T (Fig.~2a and b), we observe two lines which divide ($\Vbg$, $\Vtg$) plane into four quadrants. The vertical line corresponds to the density in the BGR ($\nbg$) equal to zero while the diagonal line follows the zero density in the TGR ($\ntg$). The crossing between these two lines at $(\Vbg^D, \Vtg^D) = (-3.2, 0)$~V indicates the charge neutrality point of the full sample. The four quadrants correspond to $nn'$, $pn$, $pp'$, and $np$ doping where the first and second letters are carrier types of the BGR and TGR respectively. Evidently, no oscillation is present at zero field. As we increase $B$ to $2$~T (Fig.~2c and d), oscillations become apparent in the $pn$ and $np$ quadrants. The curvatures of these oscillations are distinctively different from FP oscillations\cite{Young_FP_oscil} or SdH oscillations in the presence of $\pn$ junctions\cite{Williams_QHE_PN,Abanin_QHE_PN,Barbaros_QHE_PN}.

The combination of magnetic field and a $\pn$ junction leads us to consider the contribution from the snake states. Figure~2e shows a diagram of a snake state propagating along the $\pn$ interface. Here, we assume a step-like potential at the interface. With applied magnetic field, the carrier trajectory is bent into a circular orbit with cyclotron radius $\rc = \hbar \kf/ eB = \hbar \sqrt{\pi n}/eB$ where $\hbar$ is the reduced Planck's constant, $\kf$ is the Fermi wave vector, $n$ is the carrier density, and $e$ is the elementary charge. The oscillations in conductance can be understood by considering the end point at which a snake state exits the $\pn$ junction. For instance, if a snake state ends up on the collector side (Fig.~2e), this will increase the transmission probability from the injector to the collector, enhancing the conductance. However, if it exits the $\pn$ interface on the injector side (Fig.~2f), the conductance should diminish. The choice of the end point of the snake states will depend on the starting point, the length of the $\pn$ interface ($L$) and the cyclotron radii of BGR ($\rcbg$) and TGR ($\rctg$) which vary with $B$ and $n$. For example, we can move the end point of the snake state in Fig.~2e to the one in Fig.~2f by increasing $\ntg$. This enlarges the cyclotron radius in the TGR and causes the snake state to end up on the injector side instead of the collector side.

In the bipolar regime, we can determine the end point of a snake state by considering the following conditions. For $L_0 = 2N(\rcbg + \rctg) + \rcbg$ where $N$ is the integer part of the ratio $L/(2\rcbg + 2\rctg)$, a snake state will end on the collector side if $L > L_0$ and on the injector side if $L < L_0$. Using these conditions, we simulate the endpoint in the bipolar regime as a function of $\Vbg$ and $\Vtg$ where $\nbg = \cbg\Vbg^*$, $\ntg = \cbg\Vbg^* + \ctg\Vtg^*$, $C$ is the capacitive coupling, and $V^* = V-V^{\mathrm{D}}$. Figure~2g shows the result from such simulation at $B=2$~T. Here, we make the assumption, based on the semiclassical billiard model with a finite size injector\cite{Milovanovic_snake_Hall_bar}, that all snake states have the same starting point. We note that there is no fitting parameter in our simulation as all parameters ($L$, $B$, $\nbg$ and $\ntg$) can be determined experimentally. The simulation agrees relatively well with the data, especially at high density. The discrepancy at low density is likely due to disorder as cyclotron radius becomes smaller and comparable to the width of the depletion region of the $\pn$ junction (see below for detailed discussion). The contour line of a constant $\dGVtg$ is therefore described by $\rcbg + \rctg \propto \sqrt{\nbg} + \sqrt{\ntg} =$  constant, explaining the curvature of the oscillations observed in Fig.~2g. We note that, under the experiment conditions we apply, the snake states need to transmit at least $\sim$50 times across the $\pn$ interface in order to propagate from one end to the other end which is about $12.2$~$\mu$m long. This shows the robustness of the snake states compared to the skipping orbit states along a graphene edge in which the focusing peaks of the transverse magnetic focusing are greatly diminished after a few reflections of the edge\cite{Thiti_focusing,Calado_delft_focusing}. While the skipping orbits along the edge of graphene have to suffer potential fluctuation from dangling bonds, a lack of dangling bonds as well as a collimating effect of the $\pn$ junctions possibly contribute to the stability of the snake states.

%%%%%%%%%%%%%%%%%%%%%%      Figure 2     %%%%%%%%%%%%%%%%%%%%%%%%%%%%%%%%%%%%%%
%%%%%%%%%%%%%%%%%%%%%%%%%%%%%%%%%%%%%%%%%%%%%%%%%%%%%%%%%%%%%%%%%%%%%%%%%%%%%%%
\begin{figure*}
\begin{center}
\includegraphics{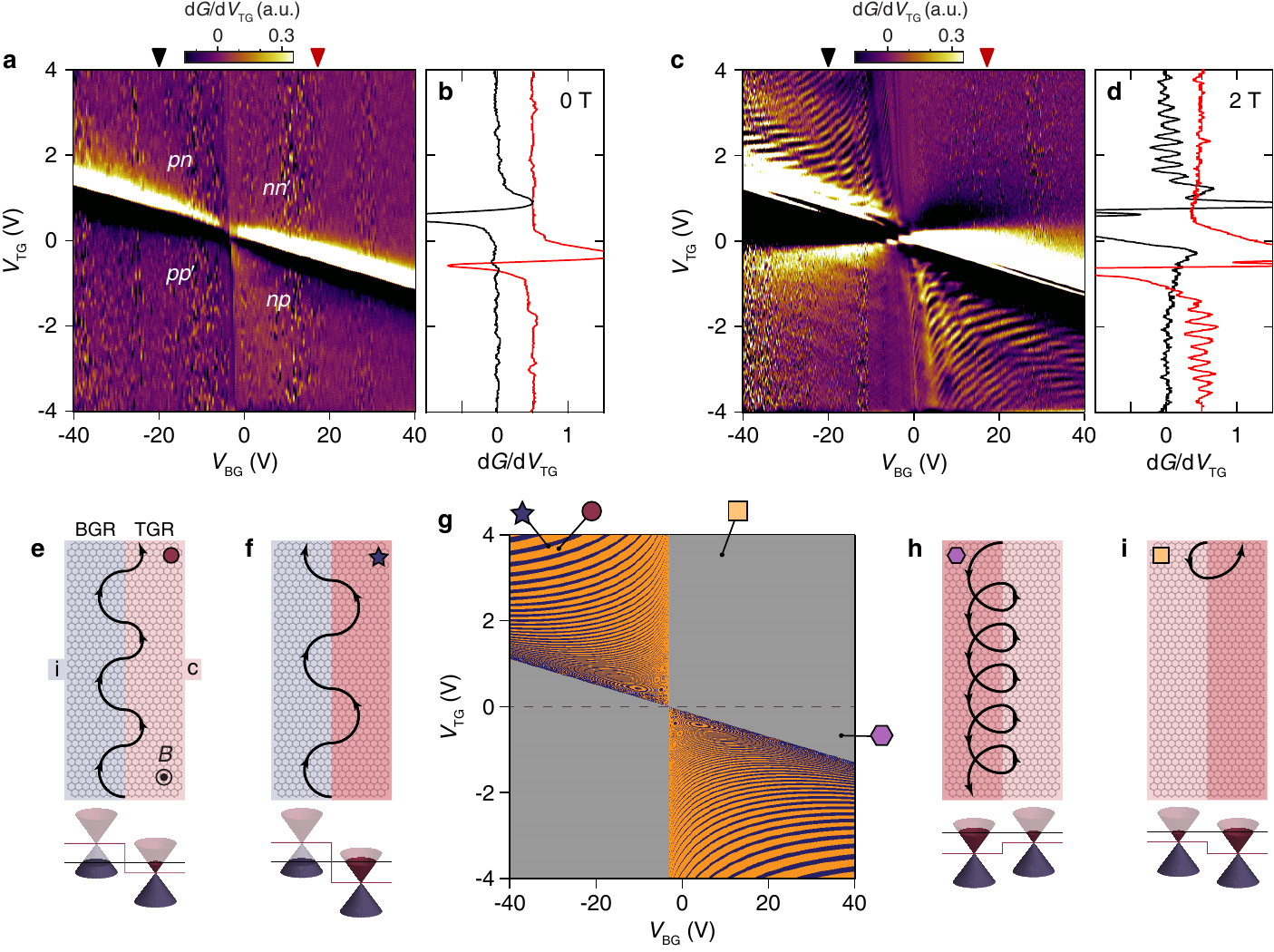}
\caption{{\bf Density dependence of the oscillating conductance and carrier trajectories along the $\pn$ interface.} ({\bf a}) $\dGVtg$ as a function of $\Vtg$ and $\Vbg$ at $0$~T. No oscillation is observed. ({\bf b}) Vertical cuts in {\bf a} at $\Vbg = 17.2$~V (red) and $-20$~V (black). ({\bf c}) $\dGVtg$ as a function of $\Vtg$ and $\Vbg$ at $2$~T. Conductance oscillations are clearly present in the bipolar regime. ({\bf d}) Vertical cuts in {\bf c} at $\Vbg = 17.2$~V (red) and $-20$~V (black). The red curves in \textbf{b} and \textbf{d} are offset for clarity. ({\bf e}, {\bf f}) Classical trajectories in the $pn$ density configuration. By tuning carrier densities, we can vary a cyclotron radius and alternate the end points of the snake states between the injector and collector sides. ({\bf g}) Simulation of the end points of snake state trajectories along the $p$-$n$ interface as a function of $\Vtg$ and $\Vbg$. When a trajectory ends on the collector (injector) side, the trajectory is represented in orange (blue). ({\bf h}, {\bf i}) The carrier trajectories in the $nn'$ density configuration. In {\bf h} ({\bf i}), the trajectories always end on an injector (collector) side.}   \label{F:Fig2}
\end{center}
\end{figure*}
%%%%%%%%%%%%%%%%%%%%%%%%%%%%%%%%%%%%%%%%%%%%%%%%%%%%%%%%%%%%%%%%%%%%%%%%%%%%%%%
%%%%%%%%%%%%%%%%%%%%%%%%%%%%%%%%%%%%%%%%%%%%%%%%%%%%%%%%%%%%%%%%%%%%%%%%%%%%%%%

In the unipolar regime, the absence of the oscillations can also be explained by similar consideration of the carrier trajectories along the interface. Since both sides of the interface have the same type of carriers, they circulate in the same direction. However, the relative magnitude of the densities on both sides produces two distinct types of trajectories.  For $\nbg > \ntg$ (Fig.~2h), we have the cyclotron radius in the BGR larger than that in the TGR and the carrier can drift along the $\pn$ interface similar to the snake states. However, the main difference is that carriers in this density configuration always end up on the injector side. Hence, the transmission probability remains unchanged as $\Vbg$ and $\Vtg$ are varied.  When $\nbg < \ntg$ (Fig.~2i), the carrier will transmit through the $\pn$ interface without drifting along the interface and always ends up on the collector side. As a result, no oscillation is achieved in both cases.

We now turn to the magnetic field dependence of the oscillating conductance. Figure~3a shows $\dGVtg$ as a function of $\Vtg$ and $B$ at $\Vbg = -20$~V. The conductance oscillations in the bipolar regime ($0.5$~V~$\leq \Vtg \leq 4$~V) are apparent from $\sim$$1$~T onward and they are much more well-behaved at low field ($< 2$~T) than at high field. To understand the observed feature, we simulate the end points of the snake states based on the aforementioned conditions.  The simulation illustrated in Fig.~3b shows that the oscillations are periodic in $B$ for a fixed $\Vtg$ with the period given by\cite{Milovanovic_snake_Hall_bar}
\begin{equation}
    \Delta B = \frac{2\hbar}{eL} (\sqrt{\pi \nbg} + \sqrt{\pi \ntg})
\end{equation}

The simulation matches the data reasonably well at low field. However, it fails to capture the onset of the conductance oscillations and aperiodicity at high field.  The onset seems to relate to the snake states undergoing a deconfinement transition below a critical field\cite{Gu_LLCollapse} $\Bc = \hbar\sqrt{\pi n}/eW$ where $W$ is half the width of the TGR. As the magnetic field is lowered, the cyclotron radius becomes larger. Consequently, once the radius is comparable to half the width of the TGR, the snake states can back scatter to the other side of the TGR edge causing the oscillations to vanish. The fit of $\Bc$ to the onset of our data yields (blue dashed lines in Fig.~3a and b) $W = 180$~nm which is roughly half the width of the TGR ($\sim$$250$ nm). Note that a similar argument can also be applied to the BGR which yields $\Bc = 0.57$~T for the 200-nm-wide half width (vertical line in Fig 3b). Moreover, the onset as a function of density at $B=1.5$~T agrees quite well with the line of constant $\rctg = 180$~nm (see Supplementary Fig. 2 and Supplementary Note 1).

The aperiodicity at high field is likely related to disorder. As the field is increased, carriers  propagate closer to the depletion region of the $\pn$ junction. In this region, the effect of the disorders becomes prominent because of the reduced screening  at low carrier density\cite{Zhang_nonlinear_screen}. The density fluctuation\cite{Xue_STMhBN,Decker_BerkeleySTMhBN} increases the roughness of the $\pn$ interface which leads to complex carrier trajectories\cite{Williams_current_guiding,Low_snake_disorder}. The interface roughness can reflect the incoming snake states due to exponential suppression of transmission rate at finite incident angle\cite{Cheianov_transmission}. Therefore, it acts as a scattering site which behaves as a new injector on the $\pn$ interface, establishing a shorter effective length ($\Leff$) the snake states needed to travel. The inset in Fig.~3b shows $\Leff$ calculated from the period in Fig.~3a (the difference in $B$ between two adjacent minima) at $\Vtg = 4$~V. At low field, the cyclotron radius is still large and most snake states can skip over scattering sites, resulting in $\Leff$ comparable to the length of our $\pn$ interface. As $B$ is increased, the snake states now with smaller cyclotron radius cannot avoid the scattering sites and, as a result, $\Leff$ decreases. The simulations in which we model the effect of disorder by putting a scattering center of finite size on the $\pn$ interface show similar increasing periods with increasing $B$ as observed in the data (see Supplementary~Fig.~3 and Supplementary~Note~2).
%%%%%%%%%%%%%%%%%%%%%%      Figure 3     %%%%%%%%%%%%%%%%%%%%%%%%%%%%%%%%%%%%%%
%%%%%%%%%%%%%%%%%%%%%%%%%%%%%%%%%%%%%%%%%%%%%%%%%%%%%%%%%%%%%%%%%%%%%%%%%%%%%%%
\begin{figure*}
\begin{center}
\includegraphics{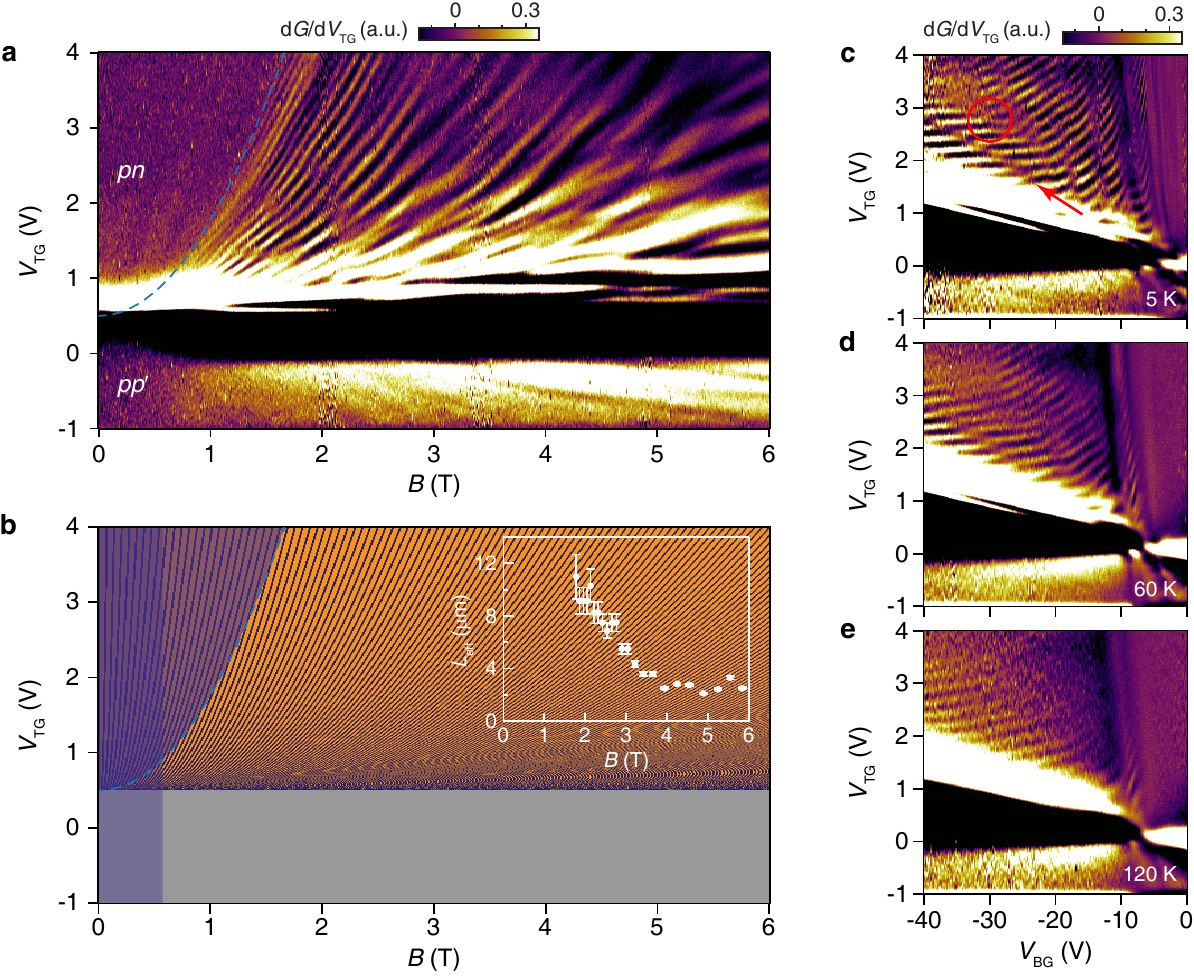}
\caption{{\bf Magnetic field and temperature dependence of the oscillating conductance.} ({\bf a}) $\dGVtg$ as a function of $\Vtg$ and $B$ at $\Vbg = -20$~V. The conductance oscillations are regular at low field and become more chaotic at high field. The blue dashed line is the critical magnetic field in the TGR. ({\bf b}) Simulation of the end points of snake state trajectories as a function of $\Vtg$ and $B$ at $\Vbg = -20$~V. The shaded regions represent the area in which magnetic field is below the critical field in the BGR (vertical slab) or TGR. Inset: The effective lengths of the $p$-$n$ interface versus $B$ at $\Vbg = -20$~V and $\Vtg = 4$~V. Error bars, calculated from Eq. 1, represent the uncertainty in determining the period of magnetic field. ({\bf c}, {\bf d}, {\bf e}) $\dGVtg$ in the $pn$ density configuration at $5$, $60$, and $120$~K respectively. The arrow and circle in {\bf c} indicate phase shift along the diagonal line and bifurcations respectively. The colour-scale bar is applied to all \textbf{c}, \textbf{d}, and \textbf{e}.}  \label{F:Fig3}
\end{center}
\end{figure*}
%%%%%%%%%%%%%%%%%%%%%%%%%%%%%%%%%%%%%%%%%%%%%%%%%%%%%%%%%%%%%%%%%%%%%%%%%%%%%%%
%%%%%%%%%%%%%%%%%%%%%%%%%%%%%%%%%%%%%%%%%%%%%%%%%%%%%%%%%%%%%%%%%%%%%%%%%%%%%%%

Finally, we comment on the temperature dependence of the conductance oscillations.  Figures~3c-e display the conductance oscillations in the $pn$ regime at $5$, $60$, and $120$~K respectively. The presence of the oscillations at such elevated temperature supports our claim that, instead of quantum interference, they are induced by snake states, which rely on ballistic transport and appearance of caustics\cite{Patel_caustics}. Simple estimation based on temperature broadening of $\kf$\cite{Young_FP_oscil,Cheianov_VeselagoLens} yields that quantum interference should be visible only up to $\sim$$10$~K. Some of the irregular features at low temperature such as phase shift (Fig.~3c, red arrow), or bifurcations (Fig.~3c, red circle) also disappear at high temperature.  This suggests that such features could be related to quantum interference between different snake states\cite{milovanovic_interplay}. More theoretical study is needed to understand the origin of these features. In addition, a new fabrication method based on the van der Waals assembly process\cite{Wang_pick_up}, which produces a large area of encapsulated graphene without bubbles and wrinkles, could potentially lower the onset of the conductance oscillations in $B$ and preserve the ballistic motion up to room temperature. This may lead to the room temperature application of the snake states such as electron waveguides\cite{Ghosh_snake_B_2} and electrically reconfigurable wiring\cite{Williams_current_guiding}.

The observation of the conductance oscillations is quite surprising given that the size of our injector is larger than $\rc$ which could have led carriers to exit the interface on both sides equally. However, under magnetic field, carriers injected into graphene will undergo a skipping orbit motion which restricts them to propagate along the edge\cite{Thiti_focusing}. The lateral extension of the skipping orbit motions will be on the order of $\rc$ which can be much smaller than the size of our injector ($\sim$300~nm). In addition, due to the exponential collimation of a graphene $\pn$ junction\cite{Cheianov_transmission,Young_FP_oscil}, only carriers which encounter the $\pn$ interface near normal incident angle will be able to pass through, owing to Klein tunneling. For carriers which collide the interface at large incident angle, they will mostly be reflected back and behave as a skipping orbit states along the $\pn$ interface. Therefore, the effect of a finite size injector is likely to introduce large conductance background and reduce amplitude of the conductance oscillations as only a small portion of electrons will turn into snake states. Moreover, simulations based on the semi-classical billiard model of a graphene $\pn$ junction have shown that the conductance oscillations still persist even for an injector whose width is wider than a cyclotron radius\cite{Milovanovic_snake_Hall_bar,Milovanovic_MEF_PN}.

Furthermore, we observe that the conductance oscillation pattern is independent of an injector location (see Supplementary~Fig.~4). A wide injector acts as a line source, instead of a point source, producing all possible skipping orbits. Hence, the distribution of electron trajectories before encountering a $\pn$ junction should be identical along graphene edge. This leads to the same initial conditions for incident electrons at the $\pn$ junction.  Therefore, an injector location does not affect the oscillation pattern as we observe.

In our analysis of the snake states so far, we have restricted ourselves to a semiclassical picture. Since most of the oscillations we observe occur at high filling factors (up to 150, see Supplementary~Fig.~5), the semiclassical picture is applicable. However, at low filling factors, evidence of the SdH oscillations, which requires a quantum approach, can be seen around ($\Vbg$, $\Vtg$) $=$ (0, 0)~V in Fig.~2c and Supplementary~Fig.~3f. These SdH oscillations disappear at high density and the oscillations due to the snake states become dominant.

In a second device we measure, there is evidence that both edge states and snake states can coexist (see Supplementary~Fig.~6). At low field, we observe the oscillations due to the snake states, similar to the first device. As we increase magnetic field, SdH oscillations and the oscillations due to the snake states are both apparent and cross each other which imply the coexistence of edge states and snake states. Further study is required to understand the interplay between the edge states and the snake states along a $\pn$ interface\cite{Zarenia_snake_dot,milovanovic_interplay} and also how energy gaps of the Landau levels affect the snake states\cite{Williams_QHE_PN,Abanin_QHE_PN,Barbaros_QHE_PN,Amet_QHE_PN}.

\subsection*{Methods}
Figure 1b shows an AFM image from one of our devices before transferring the top hBN. The position of the top gate is shown as an orange region. We first transfer graphene onto high-quality hBN\cite{Tan_graphene_hetero} and subsequently pattern it into a Hall bar shape in oxygen plasma.  Chromium and gold are used for contacts and top gate, defined by electron-beam lithography. We anneal the sample in forming gas to reduce residues from fabrication. In addition, for the device shown in Fig.~1b, we also perform mechanical cleaning\cite{Goossens_mechanical_cleaning} to further clean the device.

We measure conductance by applying a small voltage-bias excitation ($< 2$~mV) at $13$~Hz to an injector and measuring the current through a collector. The magnetic field is applied normal to graphene. All measurements were done at $5$~K unless otherwise stated. A field effect mobility of our sample is $\sim$$40,000$ cm$^{2}$V$^{-1}$s$^{-1}$.  The capacitive couplings of back gate and top gate are $\cbg = 9\times10^{-5}$~F\,m$^{-2}$ and $\ctg = 3\times10^{-3}$~F\,m$^{-2}$, determined from Shubnikov-de Haas oscillations using 4-probe measurement at high magnetic field. We observe the conductance oscillations due to snake states in two devices (see Supplementary Fig.~6 for data from a second device).

\subsection*{Acknowledgements}
We thank L. Levitov, J. Martin, and I.J. Vera Marun for discussions, A. Avsar, E.C.T. O'Farrell, and J. Wu for experimental help. B.\"{O}. acknowledges the support by the National Research Foundation, Prime Minister's Office, Singapore under its Competitive Research Programme (CRP Award No. NRF-CRP9-2011-3) and the SMF-NUS Research Horizons Award 2009-Phase II.

\subsection*{Author contributions}
T. Taychatanapat, J.Y.T., and Y.Y. fabricated the samples. T. Taychatanapat performed the experiments, analyze the data, and wrote the manuscript with input from J.Y.T and B.\"{O}. K.W. and T. Taniguchi synthesized the hBN samples. B.\"{O}. supervised the project.

\subsection*{Additional information}
Correspondence and requests for materials
should be addressed to B.\"{O}.~(email: \mbox{phyob@nus.edu.sg})

\bibliographystyle{naturemag}
%\bibliography{Snake_state}

\end{document}